\documentclass[journal,10pt]{IEEEtran}

\usepackage[utf8]{inputenc}
\usepackage{amsfonts}
\usepackage{amssymb}
\usepackage{amsmath}
\usepackage{microtype}
\usepackage{graphicx}
\usepackage{booktabs}
\usepackage{hyperref}
\usepackage{multirow}
\usepackage{footnote}
\usepackage{balance}
\usepackage{tablefootnote}
\usepackage[noadjust]{cite} % Control citation format
\usepackage{threeparttable} % for table format with note
\usepackage{gensymb} % Generic symbols for both text and math mode
\usepackage{commath} % Mathematics typesetting support
\usepackage{bm} % Access bold symbols in maths mode
\usepackage{siunitx} % A comprehensive (SI) units package
\usepackage{environ} % A new interface for environment
\usepackage{breqn} % Automatic line breaking of displayed equations
\usepackage{xcolor} % For highlight text
\usepackage{siunitx} % A comprehensive (SI) units package
\usepackage{enumitem}
\usepackage{algorithm} % for algorithm
\usepackage[noend]{algpseudocode}
\usepackage{bbold}
\usepackage[nodisplayskipstretch]{setspace}
\usepackage[skip=2pt]{caption} % example skip set to 2pt
\usepackage{subfigure}

\setlist[itemize]{nosep}
\newcommand{\highlighttext}[1] {#1}

\makesavenoteenv{table}
\makesavenoteenv{table*}
\makesavenoteenv{tabular}

\newcommand{\R}[1] {\textnormal{#1}}

\NewEnviron{NORMAL}{% 
    \scalebox{1}{$\BODY$} 
} 

\title{MMW-Carry: Enhancing Carry Object Detection through Millimeter-Wave Radar-Camera Fusion}

\author{Xiangyu Gao, Youchen Luo, Ali Alansari, Yaping Sun
\thanks{X. Gao, A. Alansari are with the Department of Electrical and Computer Engineering, University of Washington, Seattle, WA, 98195, USA. (email: xygao@uw.edu, afaa97@uw.edu). Y. Luo is with the Paul G. Allen School of Computer Science and Engineering, University of Washington, Seattle, WA, 98195, USA. (email: yluo6@uw.edu). Y.~Sun is with the Department of Broadband Communication, Peng Cheng Laboratory, Shenzhen 518000, China. (email: sunyp@pcl.ac.cn).}
\thanks{Corresponding Author: Xiangyu Gao}
}

\begin{document}
\maketitle

\begin{abstract}
This paper introduces MMW-Carry, a system designed to predict the probability of individuals carrying various objects using millimeter-wave radar signals, complemented by camera input. The primary goal of MMW-Carry is to provide a rapid and cost-effective preliminary screening solution, specifically tailored for non-super-sensitive scenarios. Overall, MMW-Carry achieves significant advancements in two crucial aspects. Firstly, it addresses localization challenges in complex indoor environments caused by multi-path reflections, enhancing the system's overall robustness. This is accomplished by the integration of camera-based human detection, tracking, and the radar-camera plane transformation for obtaining subjects' spatial occupancy region, followed by a zooming-in operation on the radar images. Secondly, the system performance is elevated by leveraging long-term observation of a subject. This is realized through the intelligent fusion of neural network results from multiple different-view radar images of an in-track moving subject and their carried objects, facilitated by a proposed knowledge-transfer module. Our experiment results demonstrate that MMW-Carry detects objects with an average error rate of 25.22\% false positives and a 21.71\% missing rate for individuals moving randomly in a large indoor space, carrying the common-in-everyday-life objects, both in open carry or concealed ways. These findings affirm MMW-Carry's potential to extend its capabilities to detect a broader range of objects for diverse applications.
\end{abstract}


\maketitle

\section{Introduction}
The escalating frequency of active shooting incidents in recent years \cite{1406480} has intensified concerns, underscoring the need for robust detection capabilities for potential threats such as firearms (e.g., shotguns, rifles) in contemporary intelligent surveillance systems. Various sensing modalities, including active millimeter-wave (MMWave) imaging \cite{TSA}, backscatter X-ray \cite{Roomi2012DETECTIONOC}, infrared \cite{817171}, Terahertz \cite{4682606}, and cameras, have been employed to address this challenge. However, existing techniques face limitations. 1) Ineffectiveness under specific scenarios: Visual cameras are hindered by concealed coverings. Infrared imaging struggles to differentiate objects with temperatures similar to the human body \cite{817171}. 2) Time and effort-consuming: Super-resolution imaging in active or backscatter sensing often necessitates a fixed posture from the subject and involves prolonged processing times for accurate results \cite{identf2012}. 

We propose \textit{``MMW-Carry"}, an economical system designed for efficient object detection through the integration of MMWave radar and a visual camera. The primary focus of this system is on cost-effective preliminary screening, specifically geared towards \textit{triggering alarms for scenarios that do not require super-sensitive detection}. Examples include less-intrusive inspections of objects for taxi passengers and building users. Due to resolution constraints, we redefine the problem as a non-conventional object detection task \cite{ramp}. In this context, the goal is not to identify the bounding box or precise location of each object but rather to predict its existence probability. To address experimental challenges related to real firearms, we narrow down the problem scope to the detection of three representative objects—laptop, phone, and knife. These items are chosen for their commonality in daily life, low radar cross-section (RCS), and shape variations, making them suitable for verifying the system's feasibility. Notably, this poses a more challenging task compared to the singular detection of firearms.

\subsection{Prior Art}
Many prior works \cite{gao2022learning, 5666180, 11198926, 5530374, 8650148} center around the concept of ``detection by imaging," involving the generation of spatial 2-dimensional (2D) or 3-dimensional (3D) object images using radar before the actual detection \cite{1406480}. Common imaging techniques include MMWave antenna scanning \cite{4682606, 5666180, 11198926, identf2012, 8650148}, synthetic aperture radar (SAR) \cite{5530374, 10.1145/3447993.3483258, gao2021mimosar}, and multiple-input and multiple-output (MIMO) processing \cite{gao2021perception, gao2019experiments, ti_mimo}. MMWave antenna scanning and SAR can yield high-resolution, vivid images but come with the trade-off of \textit{significant time and power consumption}. Moreover, the imaged subject and their carried objects must remain stationary to facilitate coherent processing. MIMO processing, while avoiding the stationary constraint, operates swiftly. However, its formed aperture is typically limited by the number of antennas, resulting in image resolution that doesn't match the first two methods \cite{gao2021mimosar}. In addition, for gait-based anomaly detection, micro-Doppler and range-Doppler images are generated and processed by CNN models and tree-based classifiers \cite{gao2019experiments, 8519763, 8536660, 10188595}.

When having a high-resolution image of the subject and their carried objects, traditional algorithms for identifying the location of specific objects (e.g., weapons) often rely on detecting contours or edges through model-based methods such as binary fitting \cite{identf2012}, expectation-maximization \cite{5666180}, and Gaussian mixture models \cite{8650148}. In addition to these model-based algorithms, an increasing number of model-free deep learning (DL) approaches have emerged for computer vision tasks. Several prior works explore the application of DL methods for weapon detection in electromagnetic signal imaging. Region-proposal-based models, such as Faster R-CNN \cite{8628238}, YOLO \cite{9269991}, and SSD \cite{9353483}, have been employed to accurately localize weapons and determine their classes. Classic CNN models, including ResNet \cite{gao2022learning} and VGG \cite{9211448}, have also been adopted to predict anomaly probabilities.

For lower-resolution images generated by MIMO, clear contours and edges for the subject, let alone smaller carried objects, are often lacking. Consequently, edge extraction-based methods, as mentioned earlier, are unsuitable. However, we have observed that different metallic objects exhibit distinct reflection patterns in radar-received signals, primarily influenced by their RCS size. Previous work also suggests the potential use of DL methods to predict the existence probability of various objects based on their unique reflection patterns in radar 3D range-azimuth-elevation images of human subjects \cite{gao2022learning}. These images are generated by a low-cost \SI{77}{GHz} MIMO radar, providing fast and single-shot imaging capabilities. Nonetheless, several challenges were left unaddressed in the previous work. Firstly, accurately locating multiple subjects in radar images within a complex indoor environment is challenging due to multipath reflections \cite{equipment2018user}. Additional false objects can occur when furniture or obstacles generate multipath reflections \cite{richards2014fundamentals}, a challenge not fully mitigated by current radar-based peak detection methods such as Constant False Alarm Rate (CFAR) \cite{nitzberg1972constant, gao2022learning}. Secondly, existing carry object detection systems are still constrained by the low resolutions of MIMO images. Approaches like using a single-frame MIMO image or voting on the results of multiple frames \cite{gao2022learning} are insufficient for effectively enhancing system performance. A more effective method of utilizing MIMO radar data or outputs is required.

\subsection{Contributions}
We expand upon the system introduced in \cite{gao2022learning} to address the aforementioned challenges while retaining the advantages of low-cost, fast imaging, and efficient object detection. Firstly, to enhance the system's robustness in complex scenarios, we integrate cameras into our setup. Cameras detect humans in the images, and the bounding box detections for each subject are subsequently smoothed and corrected using the Kalman filter. Following this, we perform radar-camera plane transformation to convert the human bounding boxes from camera image coordinates to spatial occupancy regions in the radar plane. This process ensures robust and accurate localization of subjects in radar coordinates. The localization results provide insights into the zooming-in regions for each individual in the radar image before proceeding with further detection.

Secondly, we aim to enhance carry object detection performance by leveraging long-term observations of the subject. As the subject moves, we capture multiple observations of the carried objects from different viewpoints in the radar images. These observations are obtained through the tracking results of each subject and the zoomed-in radar images. Various methods exist for combining or fusing multiple observations, ranging from simple deterministic approaches \cite{5509644, gao2022learning} to DL-based fusion modules \cite{ramp, equipment2018user}. For our specific task, we introduce an efficient late-fusion module called \textit{``knwlTrf"} designed to process neural network outputs frame by frame. Without introducing new DL layers, this module effectively discards predictions with low confidence, transferring essential knowledge over an extended period, thereby improving the system's performance.

The contributions of our work can be summarized as follows:
\begin{itemize}
    \item We present the MMW-Carry system, which combines MMWave radar and visual camera modalities to efficiently predict the presence of carried objects (e.g., laptops, phones, and knives) on moving subjects. This system offers a fast, cost-effective, and less intrusive solution.
    \item We propose a method that integrates camera-based detection, tracking, and cross-sensor plane transformation. This enables accurate localization and tracking of individuals in indoor environments with challenging multi-path reflections.
    \item We introduce a novel late-fusion module, \textit{knwlTrf}, designed to effectively combine multiple observations of different views of carried objects on moving subjects. This module improves system performance without introducing additional DL layers.
    \item We conduct extensive experiments and analyses to validate the system's performance across various scenarios and configurations. The results demonstrate the effectiveness of the proposed system in diverse real-world situations.
\end{itemize}

The paper is organized as follows: Section~II introduces the preliminaries for MMWave radar signal processing. Section III provides a detailed explanation of the system design. In Section IV, the implementation steps of the experiments are explained. Section V focuses on experimental results and analysis. Section VI concludes the paper. 

\section{Background and Preliminaries}
\begin{figure}
\centering
\includegraphics[width=0.48\textwidth]{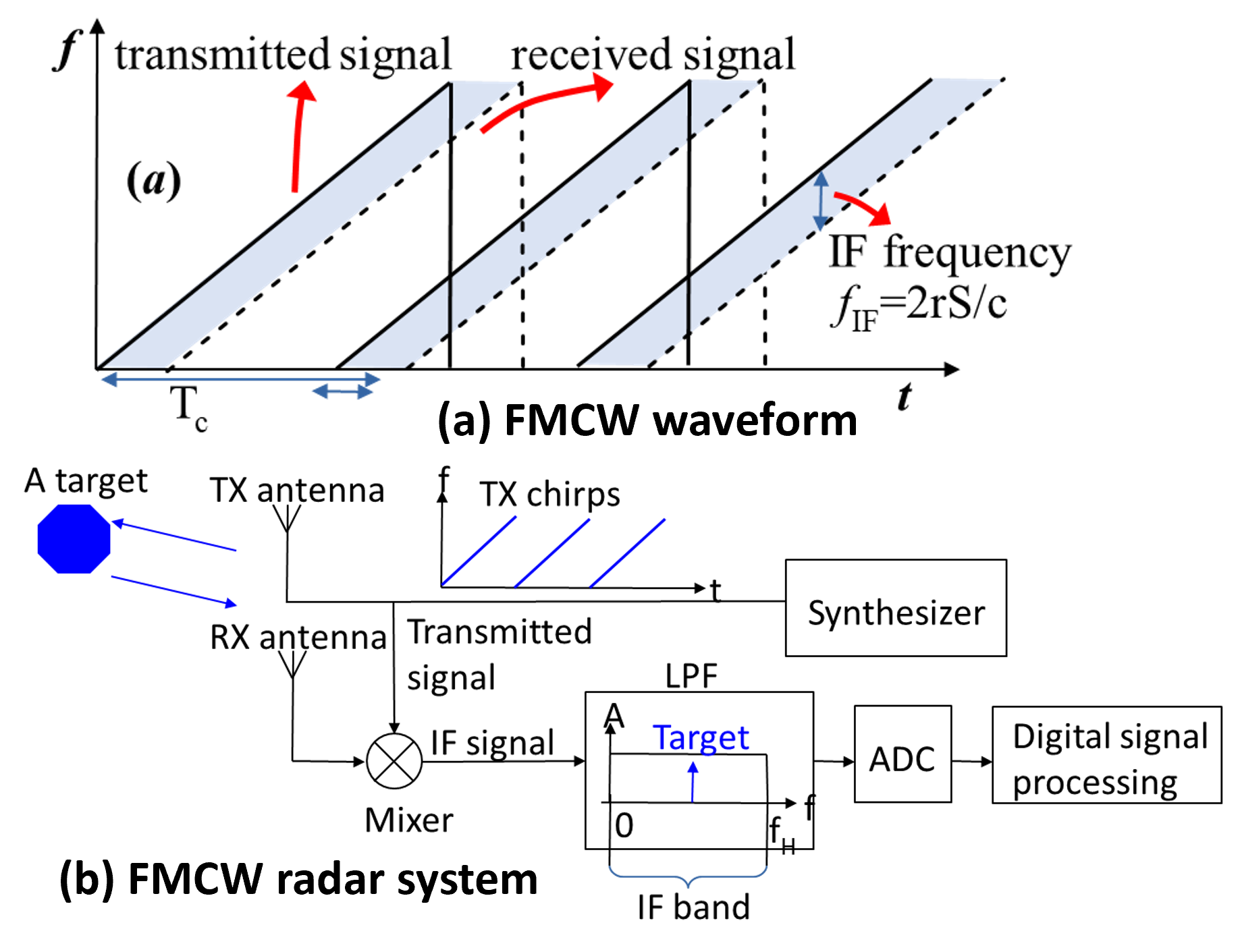}
\caption{(a) FMCW transmitted and received signal example, (b) FMCW radar system block diagram.}
\label{fig:fmcw}
\end{figure}

% \subsection{MMWave FMCW Sensing}
One frequently employed MMWave sensing device is the frequency-modulated-continuous wave (FMCW) radar, operating within the \SI{60}{GHz} or \SI{77}{GHz} bands \cite{gao2019experiments,ti_casd}. This radar system transmits a periodic wideband linear frequency-modulated signal, as illustrated in Fig.~\ref{fig:fmcw}(a), and subsequently captures the time-delayed version of the signal reflected from targets. Demodulation occurs through the multiplication of the received signal with the transmitted signal in the mixer, generating an intermediate frequency (IF) signal, as shown in Fig.~\ref{fig:fmcw}(b). The target's range and angle can then be deduced from this IF signal \cite{richards2014fundamentals, gao2019experiments}.

\subsection{Range and Angle Estimation \label{sec:theory}}
\subsubsection{Range Estimation} 
In Fig.~\ref{fig:fmcw}(a), the received signal exhibits the same slope as the transmitted signal. For a single target, a constant frequency difference between the two signals arises, resulting in an IF signal with a corresponding tone. The tone frequency linearly relates to the target range $r$ through $f_{\R{IF}} = 2rS/c$, where $c$ is the speed of light and $S$ is the sweeping slope. When multiple targets are present, the IF signal comprises a combination of multiple tones. Fig.~\ref{fig:fmcw}(b) illustrates the real part of the complex IF signal for two targets positioned at \SI{5}{m} and \SI{10}{m}. Utilizing a cost-efficient Fast Fourier Transform (FFT) facilitates the analysis of the frequency components of the IF signal \cite{gao2023static}, enabling the determination of target ranges, as depicted in Fig.~\ref{fig:fmcw}(c).

\subsubsection{Direction of Angle (DoA) Estimation} 
Angle estimation is achieved by processing the signal at a receiver array comprised of multiple elements. For a target situated at the far field and at an angle $\theta$, the resulting steering vector for a uniform linear array is given by $[1, e^{-j2\pi d\sin{\theta}/\lambda}, \ldots, e^{-j2\pi (N_{\R{Rx}}-1)d\sin{\theta}/\lambda}]^{\R{T}}$ \cite{526899}, where $d$ represents the inter-element distance, and $\lambda$ denotes the signal wavelength. Solving for $\theta$ from the array received signal is a well-explored problem \cite{CHUNG2014599}. Numerous Direction of Arrival (DoA) estimation algorithms, such as FFT-based beamformers, MVDR \cite{190331}, and MUSIC \cite{1143830}, among others, have been proposed. Notably, the FFT method, which doesn't require a priori knowledge of the number of targets, efficiently extracts the embedded phase shift $e^{-j2\pi d\sin{\theta}/\lambda}$ to rapidly resolve arrival angles $\theta$ \cite{gao2019experiments}.

\begin{figure}
\centering
\includegraphics[width=0.49\textwidth]{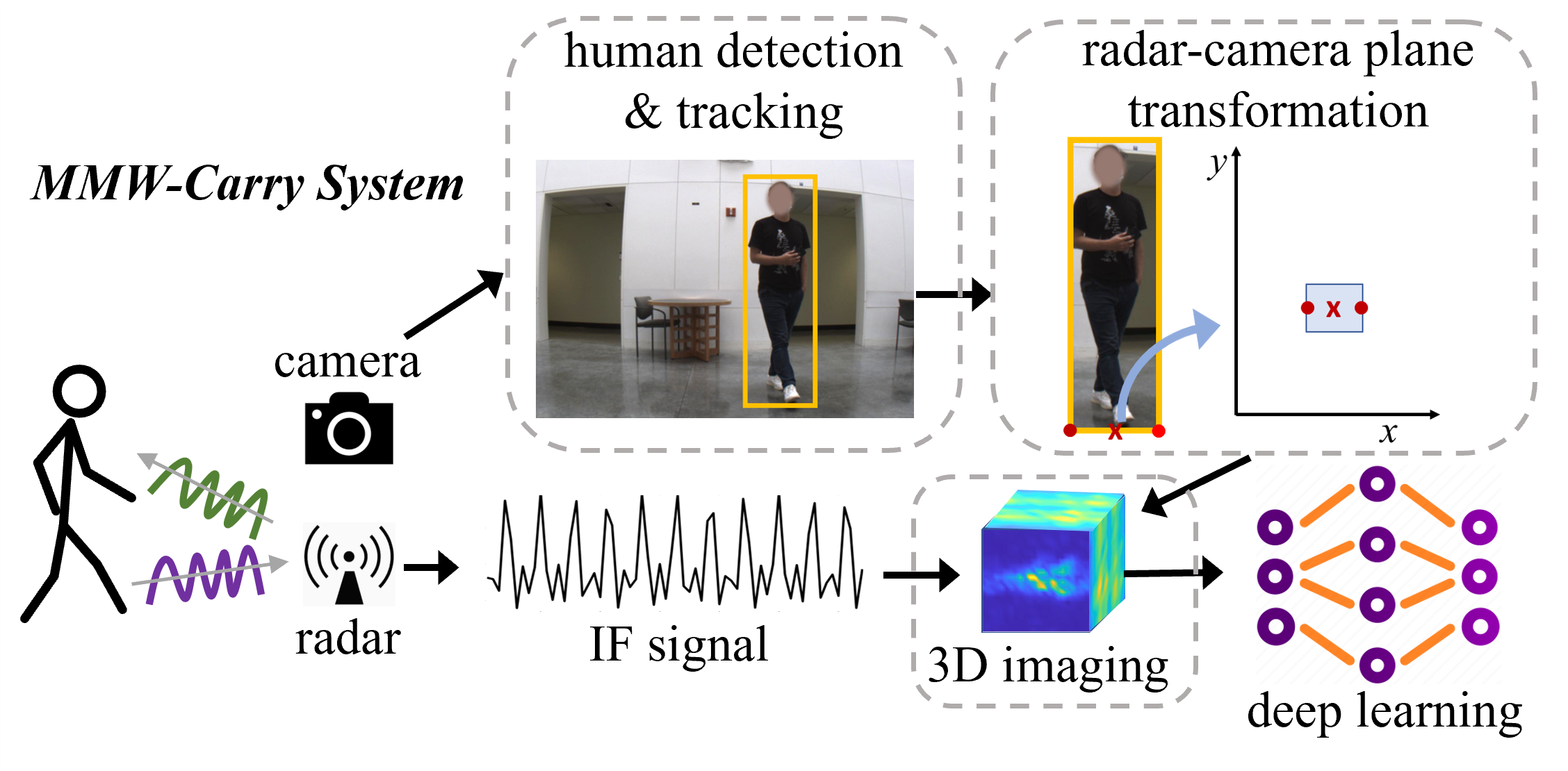}
\caption{Overview of MMW-Carry system}
\label{fig:system}
\end{figure}

\section{System Design}
We introduce the MMW-Carry system designed for detecting carried objects on individuals through a combination of MMWave radar signals and camera tracking. The overall system design, depicted in Fig.~\ref{fig:system}, comprises several key components:
\begin{itemize}
\item A human detection and tracking subsystem that generates bounding boxes of individuals in the camera images.
\item A plane transformation subsystem that projects the obtained bounding boxes onto the radar plane, defining the spatially occupied regions of the tracked human subjects.
\item A radar 3D imaging subsystem that utilizes IF signals as input, producing 3D range-azimuth-elevation imaging results for each designated human subject region.
\item A neural network model that predicts the existence probability of objects based on the radar image input. Additionally, a post-processing module refines the network output from various observations, enhancing overall system performance.
\end{itemize}

\subsection{Human Detection and Tracking}
The MMW-Carry system initiates by detecting and tracking human subjects in camera images, facilitating the subsequent localization in synchronized radar images. The integration of the camera serves to mitigate interference or multi-path reflections within complex indoor environments, thus reducing the likelihood of false or inaccurate human localization. The detailed workflow is outlined below. 

Initially, human detection employs a discriminatively trained deformable part model \cite{lsvm-pami}, which identifies individuals through a mixture of multi-scale body part models. The detection results may exhibit issues, such as extra bounding boxes on the floor due to reflections, low confidence levels, or missing bounding boxes when parts of the human body are obscured in the image due to close range or limited camera field-of-view (FoV). To address these challenges, a Kalman filter-based tracking approach \cite{kalman, 9341164} is implemented, leveraging the obtained bounding boxes.

% After implementation, a noticeable performance drop was observed in reported bounding box confidence when part of the human body is obscured from the image due to the close range and limited camera FoV. 

% After implementation, a noticeable drop was observed in reported bounding box confidence when part of the human body is obscured from the image due to the close range and limited camera FoV. 
%This issue has a tendency of causing detected bounding boxes to have a confidence below a set threshold that is appropriate for when the full body is visible.% 
% To compensate for this issue, the confidence filtering threshold is lowered when the bounding box size is relatively large to the image frame. This helps prevent the algorithm from discarding bounding boxes for human subjects that are close to the camera.

We represent the state of each subject with a tuple of 6 variables: $\left(u, v, w, l, d_{u}, d_{v}\right)$, where $(u, v)$ denote the center position of the bounding box, $(w, l)$ denote the width and length, and $\left(d_{u}, d_{v}\right)$ denote the change of $(u, v)$ from the previous frame to the current frame. To track these states, we employ a Kalman filter \cite{9341164} utilizing the constant velocity model. The tracking algorithm operates in a frame-by-frame manner, processing bounding boxes to generate tracklets. This involves updating tracklets through association and recalculation of states within the Kalman filter. The processing for each frame comprises three stages: predict, associate, and update, ensuring effective tracking of subjects over consecutive frames.
\begin{itemize}
\item \textbf{Prediction}: In the first stage, predicted bounding boxes $(\hat{u}, \hat{v}, \hat{w}, \hat{l})$ are generated for each active tracklet based on historical states using a linear relationship, i.e., $\hat{u}_{t+1} = u_t + d_{u_t}$, $\hat{v}_{t+1} = v_t + d_{v_t}$, $\hat{w}_{t+1} = w_t$, $\hat{l}_{t+1} = l_t$.
% Tracklets that have dropped more than 40 frames consecutively are considered inactive and are excluded from this step.
\item \textbf{Association}: In the second stage, detection bounding boxes are loaded, and associations with the predicted bounding boxes of active tracklets are made based on their intersection over union (IoU) values. To achieve this, a filter is applied to remove bounding boxes with significant overlaps and smaller confidence, followed by the Jonker-Volgenant algorithm \cite{Jonker2005ASA} (an $\mathcal{O}(n^3)$ alternative to the Hungarian algorithm) for association. Since the number of tracklets is not always equal to the number of detections, additional placeholder detections or tracklets may be added before running the association.
\item \textbf{Update}: In the third stage, given the matched pairs of detections and predictions, the state of each tracklet is updated to account for the uncertainty of state prediction in the Kalman filter \cite{kalman}. Additionally, we adopt the birth-and-death memory module \cite{9341164} to control the initialization and termination of tracklets. Specifically, a new tracklet is initialized and added to the active list after having matches for 20 consecutive frames, and an inactive tracklet is terminated after 40 consecutive unmatched frames.
\end{itemize}

\begin{figure}
 \includegraphics[width=0.48\textwidth]{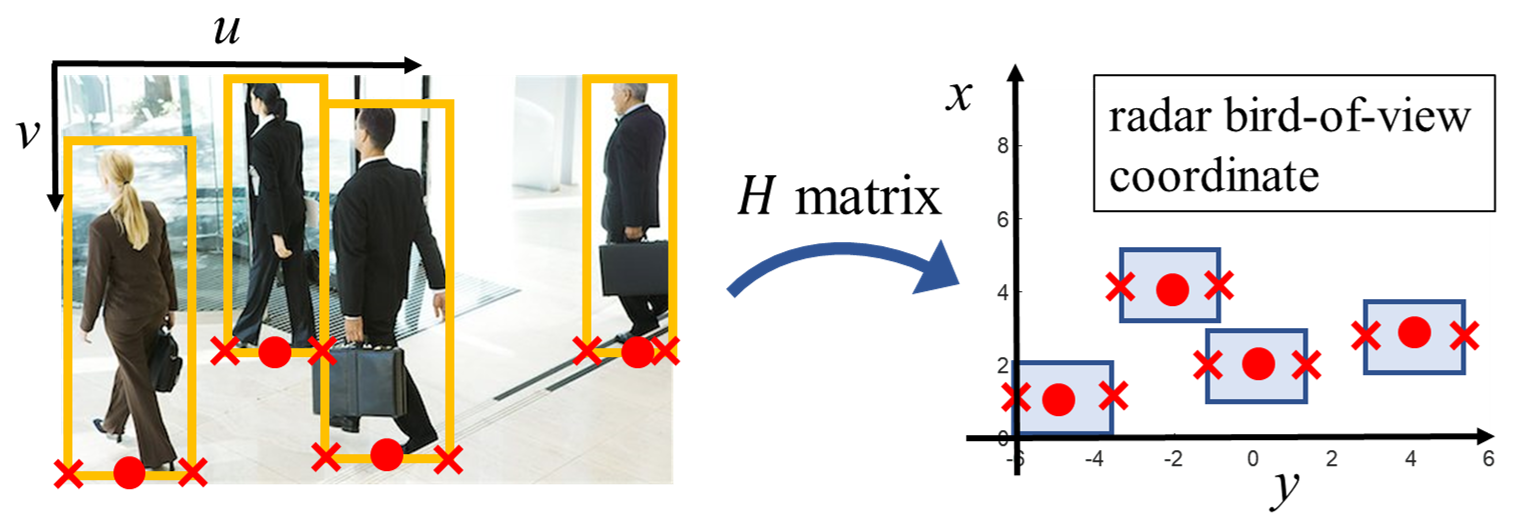}
  \caption{Transformation from the camera image plane to the radar Cartesian plane.}
  \label{fig:Transformation}
\end{figure}

\subsection{Radar-Camera Plane Transformation}
Utilizing the bounding boxes obtained from the images, we endeavored to determine the corresponding spatial locations of individuals in the radar plane. Given that the ground plane remains constant across all images (i.e., the testbed was positioned at a fixed height with a flat surface), it becomes feasible to establish a transformation that maps points $(u, v)$ on camera plan to their coordinates $(x, y, z)$ in radar plane. 

\highlighttext{Using homogeneous coordinates, we can first describe the equation of transformation between $\left(x, y, z, 1\right)$ and $(u, v, 1)$ as follows \cite{1334537}.}
\begin{equation}
\omega\left[\begin{array}{c}
u \\
v \\
1
\end{array}\right]=\mathbf{P}\left[\begin{array}{c}
x_r \\
y_r \\
z_r \\
1
\end{array}\right], \quad \mathbf{P}=\mathbf{A}\left[\begin{array}{l|l}
\mathbf{R} & \mathbf{t}
\end{array}\right]
\end{equation}
\highlighttext{\noindent where the $3 \times 3$ matrix $\mathbf{R}$ and the $3 \times 1$ vector $\mathbf{t}$ denote, respectively, the rotation and translation between the sensors' coordinates; the $3 \times 3$ matrix $\mathbf{A}$ denotes intrinsic camera parameters, and the $\omega$ is a constant. Considering that all radar data come from somewhere on the radar plane $\left(z=0\right)$ and moving $\mathbf{P}$ to the other side of the equation, we get}
\begin{equation}
\label{eq:trans}
\left[\begin{array}{c}
x \\
y \\
1
\end{array}\right]=\mathbf{H}\left[\begin{array}{l}
u \\
v \\
1
\end{array}\right],
\end{equation}
\highlighttext{\noindent where $\mathbf{H}$ is the $3 \times 3$ homography matrix. By estimating the $\mathbf{H}$, the transformation between the radar plane and the image plane is determined without solving $\mathbf{R}, \mathbf{t}$, $\omega$, and $\mathbf{A}$.}

\highlighttext{According to \cite{8581329}, we use more than four data sets of $(u, v)$ and $\left(x, y\right)$ for estimating the $\mathbf{H}$ with a singular-value-decomposition (SVD)-based constrained optimization.} Fig.~\ref{fig:Transformation} illustrates a generic example where the ground center of each person (depicted as red dots) is mapped to their respective radar coordinates using Eq.~\eqref{eq:trans} and the calculated $\mathbf{H}$. To further estimate the occupancy region of a person (shown as the blue rectangle), we assume the projected ground center as the spatial center. Two bottom vertices of the image bounding box (red crosses) are projected into radar coordinates, allowing the calculation of their distance along the $x$-direction. This distance is then used as the width, and a predefined length (sufficient to encompass a person with various poses) is employed. It's worth noting that all these processes involve straightforward and time-efficient linear operations.

In practical scenarios, there might be a projection offset due to errors in the generated bounding boxes, particularly on the bottom, which may not accurately represent the true spatial center of the human in the radar image. To address this, we adjust the estimated center in the radar image to a nearby peak with a similar azimuth angle and the strongest amplitude. We then calculate the average moving offset, which is subsequently compensated in the projection of future frames.

\subsection{Radar 3D Imaging}
\begin{figure}
\centering
\includegraphics[width=0.49\textwidth]{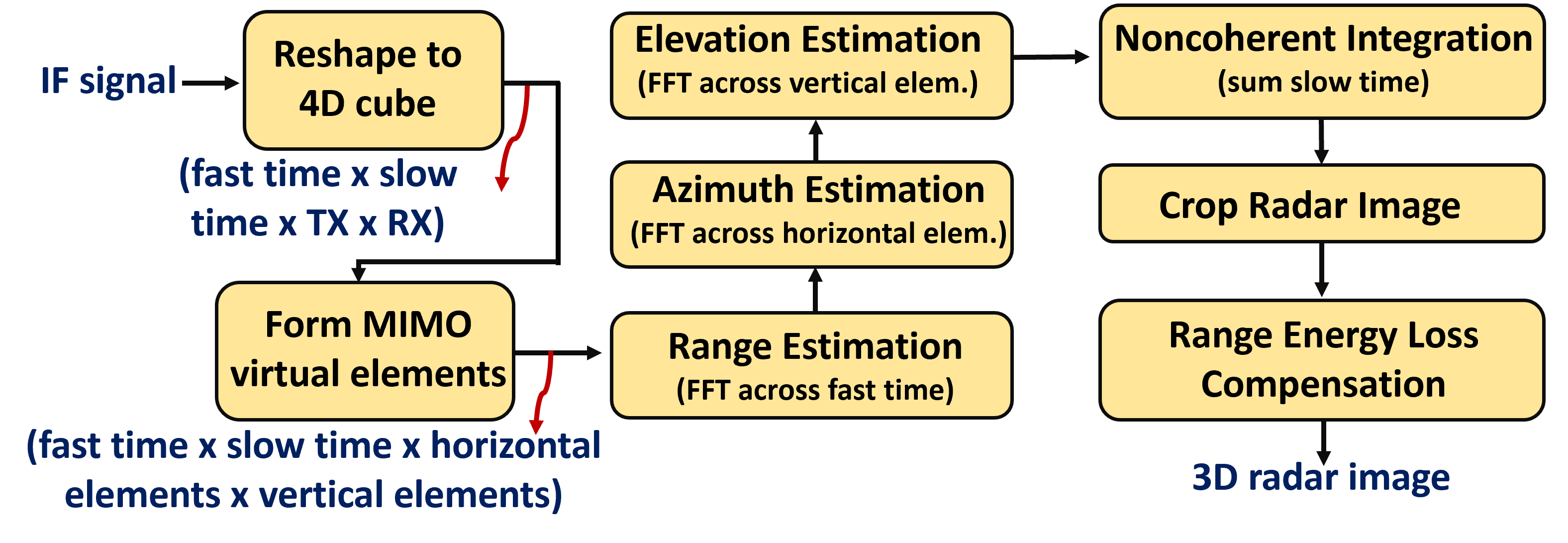}
\caption{Signal processing pipeline for 3D radar imaging from IF signal.}
\label{fig:sp}
\end{figure}

\begin{figure}
\centering
\includegraphics[width=0.45\textwidth]{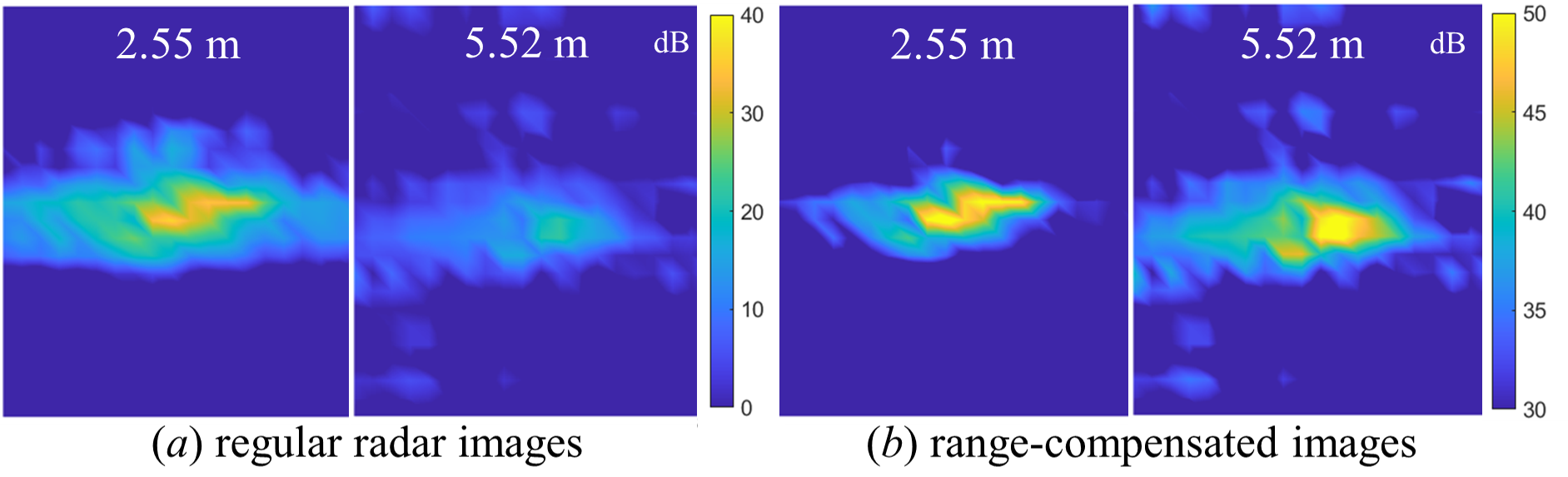}
\caption{Radar range-angle images (a) without and (b) with energy compensation for different ranges, \SI{2.55}{m} and \SI{5.52}{m}.}
\label{fig:range_comp}
\end{figure}

\begin{figure*}[t]
\centering
\includegraphics[width=0.99\textwidth]{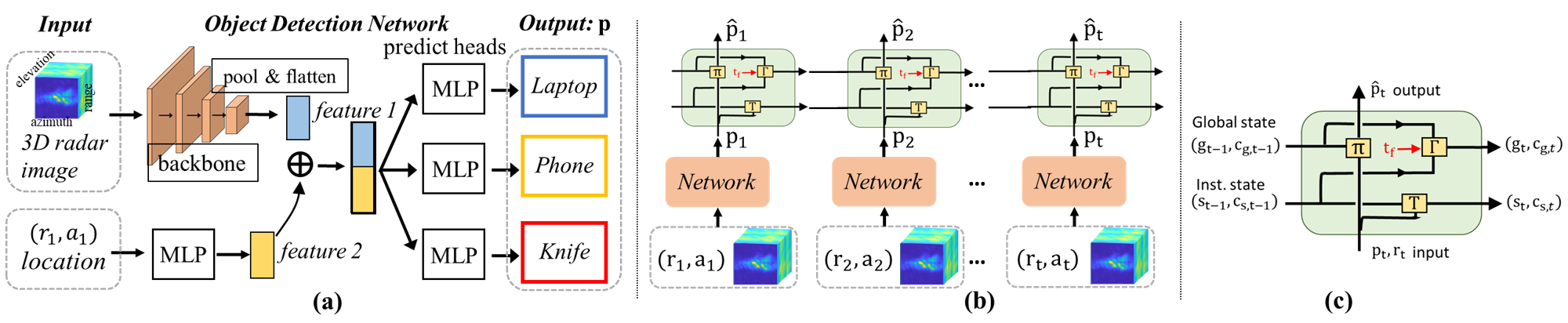}
\caption{(a) Neural network model (b) Multi-observation post-processing (c) Knowledge-transfer-based decision module.}
\label{fig:radarnet}
\end{figure*}

\subsubsection{3D Radar Imaging} 
To obtain a comprehensive 3D representation of each individual and their carried objects, we implemented range-azimuth-elevation estimation, generating a 3D spectrum from the detected IF signal \cite{gao2022learning}. For each occupied region, transformed from the camera bounding box, an image cube is extracted from the spectrum, with subsequent energy loss compensation. The signal processing workflow details are shown in Fig.~\ref{fig:sp} and outlined below.

\highlighttext{First, we reshape the sampled IF signal into a 4D cube with fast time, slow time, TX, and RX dimensions. When the transmitters sequentially transmit chirp signals ensuring orthogonality (commonly referred to as time-division multiplexing MIMO) \cite{gao2021perception}, MIMO theory can be used to efficiently enhance angular resolution by creating a virtual array that spans a larger aperture. Under the far-field assumption, with the transmitter and receiver locations denoted as $\mathbf{L}_t$ and $\mathbf{L}_r$, respectively, the virtual array formed by MIMO is situated at the spatial convolution of $\mathbf{L}_t$ and $\mathbf{L}_r$ \cite{li2007mimo}. By adhering to MIMO theory, we generate a new 4D cube, with the last two dimensions housing the horizontal and vertical elements of the resultant 2D virtual array. Then, three FFTs are conducted on the 4D cube across its fast-time, horizontal elements, and vertical elements of the virtual array. The FFT across the fast-time dimension extracts dominant frequencies of the IF signal for range estimation, while the other two FFTs across array elements serve as conventional DoA estimators. This process results in a 3D image providing a comprehensive map of the radar FoV. To enhance the quality of the 3D radar image, we make the noncoherent integration by summing the images of different slow times (chirps).}

Second, extract the radar image for each individual by cropping from the overall imaging result based on the occupancy region. The cropped elevation is fixed for all individuals to encompass most of the height of the human body. Additionally, the cropped image is padded with zeros to create a fixed-size input of $24 \times 24 \times 10$ for the neural network.

Third, compensate for the energy loss induced by range differences in the cropped radar images. According to the radar range equation \cite{doi:10.1036/0071444742}, the amplitude of the received signal is inversely proportional to the square of the range $r$, resulting in varied power values for radar images at different ranges, as depicted in Fig.~\ref{fig:range_comp}(a). To prevent signal features from being too scattered for effective learning by the network, we compensate for the range loss in signal amplitude $A$ of images by multiplying with $r^2$ \cite{doi:10.1036/0071444742, ramp}. Post-compensation, the results in Fig.~\ref{fig:range_comp}(b) demonstrate that two radar images with different center ranges are scaled to similar amplitude levels.

\subsection{3D Radar Image Object Detection}
The comprehensive architecture for object detection in 3D radar images, an extension of the work presented in \cite{gao2022learning}, comprises a backbone, prediction heads, and a knowledge-transfer-based decision module. This system takes radar images and their central locations as input, producing the existence probability for three classes of objects in each frame. The knowledge-transfer-based decision module enhances performance by transferring result knowledge across time. The details of each component are introduced below.

\subsubsection{Backbone} 
The selected backbone is a customized version of ResNet \cite{7780459}, incorporating 3D convolutions to extract crucial features from each pair of 3D radar image and its central location $(r, a)$ in range and azimuth angle \cite{gao2022learning}. As depicted in Fig.~\ref{fig:radarnet}(a), the features from the last layer undergo global average pooling for condensation, followed by concatenation with the location feature embedding acquired through a straightforward multi-layer perceptron (MLP).

\subsubsection{Prediction Head} 
Utilizing the features extracted from the image-location pair, a prediction head is designed to deduce the existence probability of an object class by processing these features. In this work, we present a combination of three prediction heads for detecting laptops, phones, and knives, respectively, as illustrated in Fig.~\ref{fig:radarnet}. Similar to the approach in \cite{https://doi.org/10.48550/arxiv.2005.12872, gao2022learning}, each prediction head is a 5-layer MLP with \textit{Sigmoid}, producing a probability $p \in [0,1]$ for a given class. When $p$ surpasses a predefined threshold $p_{\R{thr}}$, we assume the object exists. The adoption of three independent binary prediction heads, instead of a single 3-class prediction head (e.g., with \textit{Softmax}), allows for an independent existence assumption and accommodates scenarios involving the coexistence of multiple objects or no objects at all.

\subsubsection{Knowledge-transfer-based decision}
The introduced backbone and prediction heads effectively process each radar image input, producing a prediction result. Assuming the prediction result for the carried object and the center distance of the individual is denoted as $p_t$ and $d_t$ for frame $t$, we aim to enhance decision-making by designing a module capable of aggregating results across consecutive frames. This module, termed \textit{``knowledge-transfer-based decision"} or \textit{``knwlTrf"}, addresses two critical issues with its design.

Firstly, we maintain the essential knowledge of prediction results and discard less reliable information by introducing the \textit{global state} $(g_{t}, c_{g, t})$ and \textit{instantaneous state} $(s_{t}, c_{s, t})$. Regarding the global state, we focus on knowledge transfer over an extended period, where $g_t$ signifies the global prediction confidence up to time $t$, and $c_{g, t}$ represents the count of instances where the object was predicted to exist before time $t$. As for the instantaneous state, we consider knowledge accumulation over a brief timeframe, specifically when the range $r_t$ remains in the same range bin (i.e., $r_t = r_{t-1}$). Here, $s_t$ denotes the sum of prediction confidence over the short time, and $c_{s,t}$ is a counter similar to the one used in the global state ($c_{g, t}$). The update for the instantaneous state $(s_{t}, c_{s, t})$ is defined by the function $\mathbf{T}(\cdot)$.
\begin{equation}
(s_{t}, c_{s, t}) =  \delta_{r_{t}=r_{t-1}} \cdot (s_{t-1}, c_{s, t-1}) + (p_{t}, 1),
\label{eq:s_update}
\end{equation}
\noindent where $\delta_{(\cdot)}=1$ if the statement in $(\cdot)$ is true. Eq.~\eqref{eq:s_update} simplifies the aggregation of prediction results for a subject moving within a short range (a range bin) to obtain $(s_{t}, c_{s, t})$. If the subject moves out of the current range, i.e., $r_{t} \neq r_{t-1}$, the state $(s_{t}, c_{s, t})$ is reset to $(p_t,1)$.

The update for the global state $(g_{t}, c_{g, t})$ is defined by the function $\mathbf{\Gamma}(\cdot)$ as given by Eq.~\eqref{eq:g_update}:
\begin{equation}
(g_{t}, c_{g, t}) = (g_{t-1}, c_{g, t-1}) + \delta_{r_{t} \neq r_{t-1}} \cdot (s_{t-1}, c_{s, t-1}) \cdot t_f,  \label{eq:g_update}
\end{equation}
\noindent where $t_f$ is a transfer coefficient determined by the value of $p_{eval} = \frac{s_{t-1}}{c_{s, t-1}}$. If $|p_{eval}-0.5| \ge \varepsilon$, we assign a high $t_f$ value $1$; otherwise, a small value $0$. The parameter $\varepsilon$ is chosen based on the quality of predictions $p_t$, calculated as their deviation within an update cycle. When the deviation is large, indicating dispersed predictions, we use a higher threshold value, e.g., $\varepsilon=0.1$. The global state is updated by accumulating the instantaneous state with a knowledge transfer coefficient $t_f$ when the subject moves to a new range. This approach helps retain valuable knowledge and prevents unwanted confidence from affecting the global state.

The second issue addressed is maintaining consistency for frame-by-frame decisions while issuing instantaneous negative notifications. This is accomplished by averaging the prediction input $p_{t}$ with the global state $(g_{t-1}, c_{g, t-1})$ using a defined function $\mathbf{\Pi}(\cdot)$ as expressed in Eq.~\eqref{eq:p_ouput}. The refined decision $\hat{p}_{t}$ is then outputted through this process.
\begin{align}
& \hat{p}_{t} = \frac{g_{t-1}+p_{t}}{c_{g, t-1}+1} 
\label{eq:p_ouput}
\end{align}

\subsubsection{Other Fusion Methods\label{sec:fuse_baseline}} 
The proposed knwlTrf module operates as a late-fusion technique, integrating data at the result level. For comparison, we also explore the following fusion methods: attention-based feature fusion, denoted as \textit{``feaAtt"}; feature concatenation fusion, denoted as \textit{``feaCat"}; history feature accumulation, denoted as \textit{``feaAcm"}; and another late-fusion method using result voting \cite{gao2022learning}, denoted as \textit{``resVoteShort"}.

\begin{figure}
\centering
\includegraphics[width=0.48\textwidth]{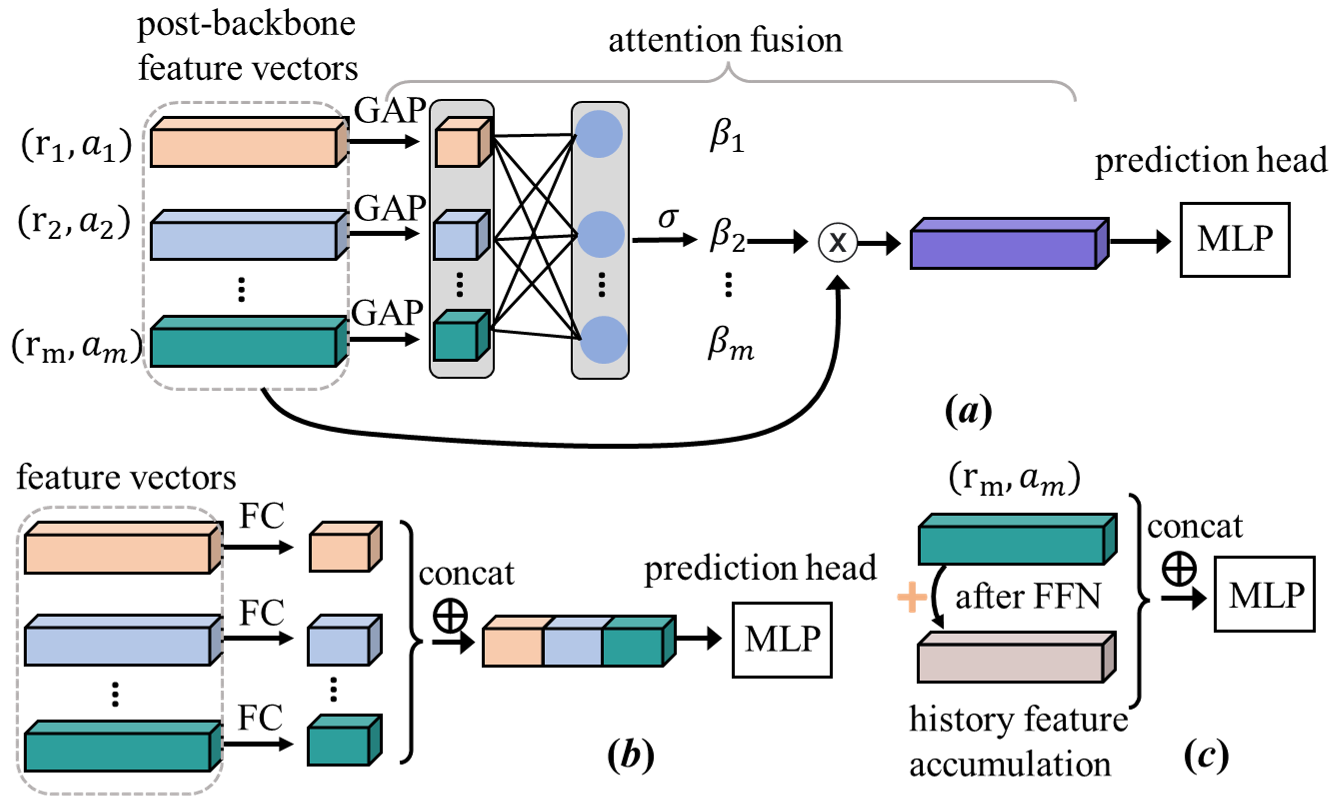}
\caption{Fusion Methods Comparison: (a) Attention-based feature fusion (\textit{feaAtt}), (b) Feature concatenation fusion (\textit{feaCat}), (c) History feature accumulation (\textit{feaAcm}).}
\label{fig:fusion}
\end{figure}

The first three methods fall under feature-level fusion, as depicted in Fig.~\ref{fig:fusion}(a)-(c), where features from multiple frames are combined and then processed by the MLP for prediction. Specifically:
\textit{feaAtt} employs a cross-attention module to predict the weight of each input feature and performs a weighted sum accordingly.
\textit{feaCat} projects feature vectors to a smaller size through a fully connected layer and concatenates all shrunken features along the channel dimension.
\textit{feaAcm} linearly accumulates all history features and utilizes the concatenation of the history feature accumulation and the current feature as input to support the prediction.
While \textit{feaAtt} and \textit{feaCat} may be limited in processing a specific number of features due to network dimensionality issues, \textit{feaAcm} accumulates all history features, offering robust prediction support.

In contrast to the feature-level fusion methods mentioned above, the \textit{resVoteShort} method makes the final decision based on the majority principle using the prediction results from the last few frames. This approach aims to smooth and reduce prediction noise. However, it can accumulate incorrect decisions if there are numerous flawed detections.

\subsubsection{Loss Function} 
The loss function for this network is the weighted sum of the Focal Losses \cite{ gao2022learning} for all the classes, same as \cite{gao2022learning}. 
% as expressed in Eq.~\eqref{eq:loss}. Focal Loss is employed to address class imbalance during training, where the number of positive (existing) objects for each binary prediction head is less than the total number of negative (not existing) objects.
% \begin{equation}
% \begin{aligned}
% &\qquad \qquad \text{Loss} = \text{FL}_{\R{laptop}} + \text{FL}_{\R{phone}} + \text{FL}_{\R{knife}}\\
% &\text{FL}(p) = - w_1 y (1-p)^\alpha \log p - w_2 (1-y) p^\alpha \log (1-p)
% \end{aligned}
% \label{eq:loss}
% \end{equation}
% \noindent where $y$ represents the ground truth, and $p$ is the predicted probability for a class. The parameters $\alpha$, $w_1$, and $w_2$ are the focusing and weight-balance parameters. In this paper, we found that $\alpha=4$ is effective for focusing on hard cases.

\begin{figure}
\centering
\includegraphics[width=0.48\textwidth]{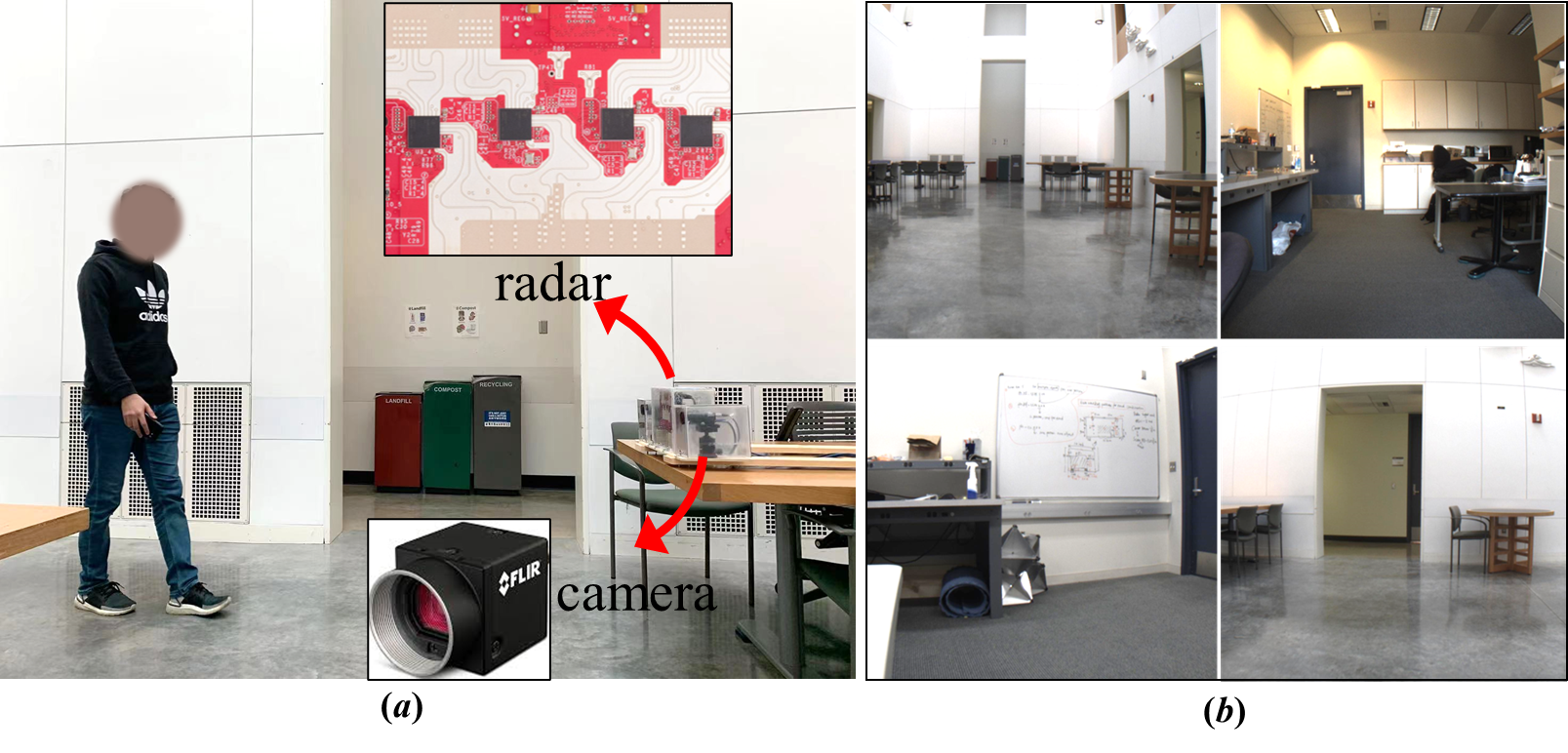}
\caption{(a) Implementation and experiment setup (b) Data collection locations.}
\label{fig:testbed}
\end{figure}

\section{Implementation}
% In this section, we describe our setup, dataset, radar-camera plane calibration, and signal processing procedure.
\subsection{Testbed Setup and Dataset}
\subsubsection{Tested}
Our testbed comprises two FLIR cameras and one  \SI{77}{GHz} MMWave radar from Texas Instruments \cite{ti_casd} as shown in Fig.~\ref{fig:testbed}(a). The setup is enclosed in a transparent box, allowing visibility while concealing the internals. \highlighttext{The configuration of this radar is presented in Table.~\ref{tab:sys_param}. Based on those parameters and \cite{gao2019experiments}, the specifications of the radar can be given out in terms of range resolution ($\frac{c}{2B}=\SI{0.06}{m}$), maximum detectable range ($\frac{f_{\R{s}} c}{2S} = \SI{15}{m}$). The radar is equipped with 12 TX antennas and 16 RX antennas. With time-division multiplexing (TDM) on TXs, it can form a large 2D virtual array with 192 elements via the spatial convolution of all TX and RX, resulting in fine azimuth resolution (\SI{1.35}{\degree}) and elevation resolution (\SI{19}{\degree}) \cite{gao2022learning}. }
% , Doppler velocity resolution ($\frac{\lambda}{2N_{\R{c}} T_{\R{c}}}=\SI{0.072}{m/s}$), and maximum operating velocity ($ \frac{\lambda}{4T_{\R{c}}} = \SI{1.80}{m/s}$).

\begin{table}
\centering
\setlength\tabcolsep{3pt} 
\caption{Radar Configuration}
\begin{tabular}{ll}
\hline  
Configuration & Value\\
\hline  
Frequency ($f_{\R{c}}$) & \SI{77}{GHz}\\ 
Sweeping Bandwidth ($B$) & \SI{2.5}{GHz}\\
Sweep slope ($S$) & \SI{79}{MHz \per\micro\second} \\
Sampling frequency ($f_{\R{s}}$) & \SI{8}{Msps}\\
Num. of chirps in one frame ($N_{\R{c}}$) & $50$ \\
Num. of samples of one chirp ($N_{\R{s}}$) & $256$ \\
Duration of chirp\footnote{$T_{\R{c}}$ is the multiplication of single chirp interval and TX number, $T_{\R{c}}=\SI{45}{us} \times 12=\SI{540}{us}$.} and frame ($T_{\R{c}}$, $T_{\R{f}}$) & \SI{540}{\micro\second}, \SI{1/30}{s} \\
\hline 
\label{tab:sys_param}
\end{tabular}

\end{table}

\subsubsection{Dataset}
We have amassed an extensive and diverse dataset encompassing synchronized camera images and MMWave radar signals. The dataset features individuals carrying a range of common objects, including phones, laptops, knives (e.g., metallic butter knives and cutting knives), and others such as keys or no object. For more specific security applications, the inclusion of additional dangerous objects like guns is a potential avenue for future work, pending permit considerations. The dataset statistics are outlined below \footnote{A portion of our dataset is accessible on IEEE Dataport \cite{begn-ye78-22}.}:

\begin{itemize}
\item \textbf{Scenarios}: Individuals walk at an average pace in typical spaces, following random paths.

\item \textbf{Scale}: The dataset encompasses 3.4 hours of data, comprising 367,482 samples of synchronized camera image frames and MMWave signals.

\item \textbf{Diversity}: Data is collected from 12 distinct locations on a university campus, including building lobbies, laboratories, offices, and open spaces (see Fig.~\ref{fig:testbed}(b)). Frames typically depict either one or two people randomly walking, each carrying one or multiple objects. Object status can be either concealed or openly carried, with concealed objects accounting for $48\%$ in regular walking scenarios. The objects vary in weight, size, shape, and carrying method. 
% To enhance dataset variability, the location and manner of object carrying were periodically altered, and individuals' walking paths were consistently randomized.
\end{itemize}

\subsection{Radar-Camera Plane Calibration}
\begin{figure}
\centering
\includegraphics[width=0.35\textwidth]{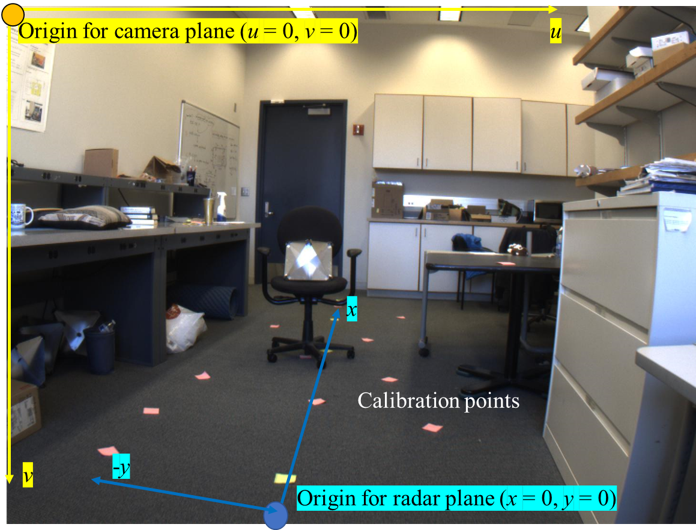}
  \caption{The camera, radar plane definition, and measurement of reflectors for calibration experiments.}
  % The colored marks on the ground were for reference and measurement of the ground center of the reflector on the chair
  \label{fig:plane}
\end{figure}

To project the camera plane onto the radar plane using Eq.~\eqref{eq:trans}, the matrix $\mathbf{H}$ is determined through the solution of equations derived from a minimum of 4 calibration points. Calibration data was collected by placing a reflector on a chair at various locations, each marked with pink stickers (see Fig.~\ref{fig:plane}). The image pixel corresponding to the sticker beneath the chair and the reflector's location in processed radar images were manually determined for calibration. Throughout the experiment, 19 data points were recorded across two rooms. Due to this limited number, cross-validation was employed for performance evaluation. In each run, 18 out of the 19 points were selected as the calibration set to solve for $\mathbf{H}$, while the remaining point served as the test set. This approach ensured that each data point had the opportunity to be evaluated. The overall results are presented in Fig.~\ref{fig:trans_testing}(a), showcasing the estimated reflector locations in the radar plane (red points) using Eq.~\eqref{eq:trans} alongside the ground truth (blue points). Statistically, the transformation exhibited relatively small errors from the ground truth, with an average of \SI{3.4}{inches}. The error is expected to decrease with the inclusion of more data points in the calibration process.

% \subsection{Signal Processing}
% The hyperparameters employed for radar signal processing are summarized as follows: Range FFT points 256, Azimuth Angle FFT points 86, and Elevation Angle FFT points 16. Hann windowing was applied to all FFTs to mitigate side-lobe levels \cite{richards2014fundamentals}. Additionally, all cropped radar images were normalized using a coefficient of $10^5$.

\section{Experiments and Evaluation}
\subsection{Overall Performance}
The overall performance of the MMW-Carry system is assessed by comparing it with baselines using the entire testing set, encompassing concealed and open carry, as well as single and multiple carried objects.

\subsubsection{Training/Testing Set} 
Approximately $70\%$ of the data was allocated for training, with the remaining $30\%$ reserved for testing. Ensuring that the testing and training data do not overlap in sequences is a crucial step. To facilitate training with a substantial amount of data, we initially start with a small subset for initialization training. The learning rate commences at $10^{-3}$ and is halved every 10 epochs. Additionally, optimization was performed using an SGD optimizer with a batch size of $256$. To strike the optimal balance in performance for the three prediction heads, we froze the backbone and fine-tuned them towards the end of the training process.

\subsubsection{Metrics} 
We employed the followings for evaluation:
\begin{itemize}
    \item \textit{False Positive Rate (FPR)}: This is defined as the fraction of falsely detected objects among all actual negative cases. It gauges the system's robustness against spurious detections.
    \item \textit{Missing Rate (MR)}: This is defined as the fraction of falsely ignored objects among all actual positive cases. It assesses the system's robustness against missed detections.
    \item \textit{Accuracy}: This is defined as the fraction of all correctly classified objects among all cases.
\end{itemize}

\subsubsection{Baselines} 
The baselines for comparison include: \textit{singleFram} that utilizes only one frame as input; \textit{resVoteShort} that combines the prediction results of multiple frames through voting; \textit{feaAcm} that accumulates history features for prediction (refer to Section \ref{sec:fuse_baseline}). These baselines were chosen for comparison as they do not require the use of delay-lag feature vectors and the training for fusing them via a neural network, as is the case with methods like \textit{feaAtt} and \textit{feaCat}. The detailed comparison with \textit{feaAtt} and \textit{feaCat} is presented in Section~\ref{sec:combines} with 10 frames of data as input.

\subsubsection{Performance} 
The performance of the proposed system with knwlTrf and its comparisons are summarized in Table~\ref{tab:walk_Resl}. In the context of regular walking scenarios, the system's performance demonstrates improvement when temporal information (i.e., multiple observations) is considered. Compared to the singleFrame method, the knwlTrf method exhibits a reduction of approximately 10\% in both false positive and missing rates, coupled with an enhancement of around 10\% in accuracy. In comparison to the resVoteShort and feaAcm methods, the knwlTrf method holds a slight performance advantage across all metrics. This superiority arises from its ability to mitigate the impact of erroneous predictions while preserving accurate ones through the proposed knowledge transfer mechanism.

\begin{table}
    \centering
    \setlength\tabcolsep{3pt} 
     \caption{Overall evaluation for collected test set.}
    \begin{tabular}{cccc}
        \hline
         Methods & FPR & MR & Accuracy\\
        \hline
         singleFrame & 0.3535 & 0.3145 & 0.6625\\
         resVoteShort & 0.3030 & 0.2716 & 0.7110\\
         feaAcm & 0.2563 & 0.2400 & 0.7490\\
         knwlTrf (\textbf{ours}) & \textbf{0.2522} & \textbf{0.2171} & \textbf{0.7619}\\
        %  track-result_drop & \textbf{0.2522} & \textbf{0.2171} & \textbf{0.7619}\\
         \hline
    \end{tabular}
    \label{tab:walk_Resl}
\end{table}

\begin{figure}
\centering
\subfigure[]{\includegraphics[width=0.24\textwidth]{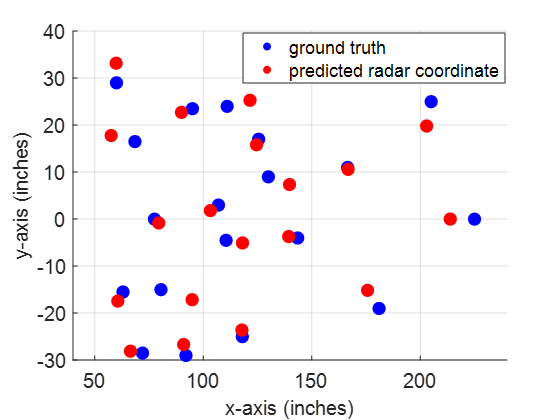}}
\subfigure[]{\includegraphics[width=0.24\textwidth]{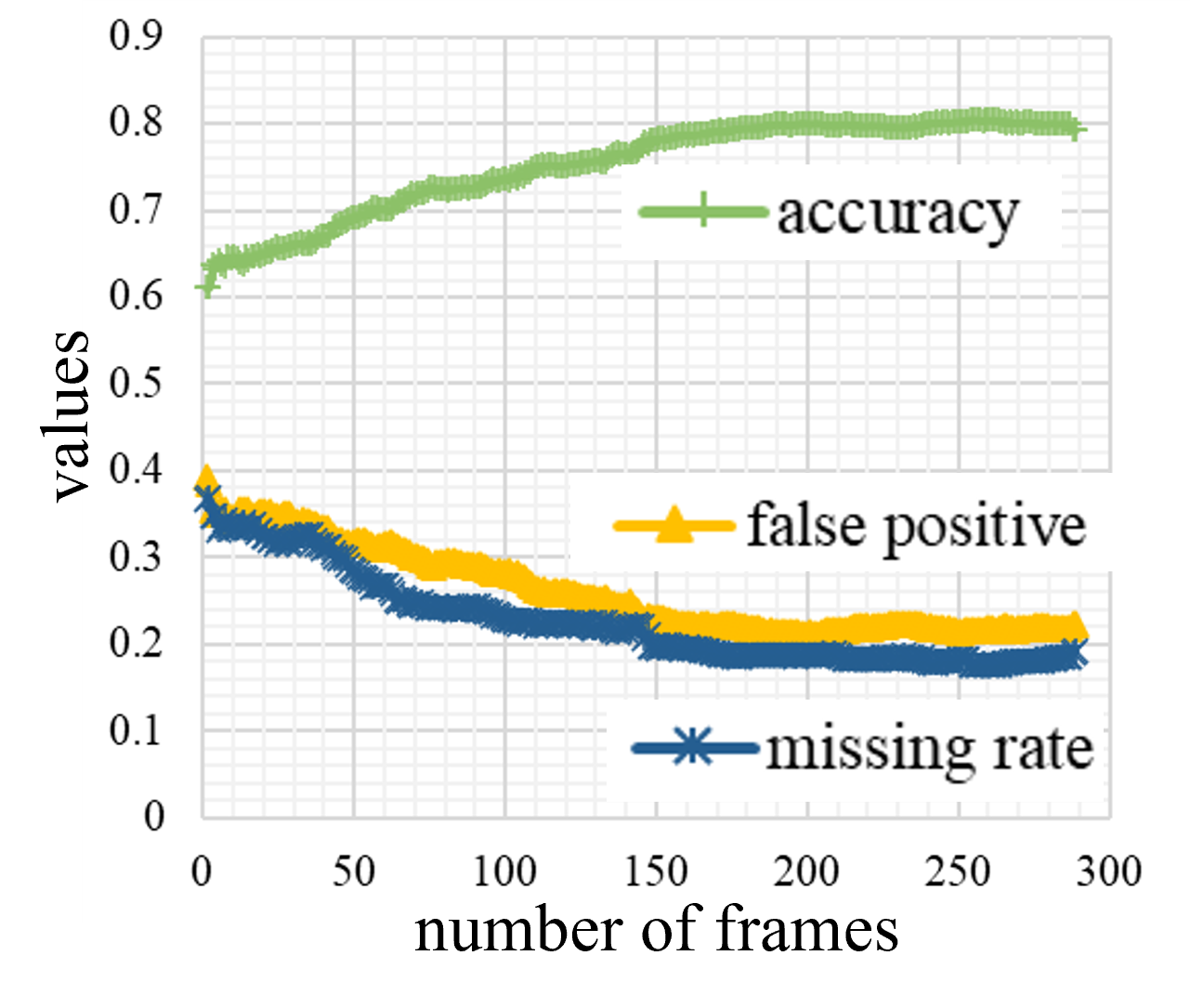}}
\caption{(a) Cross-validation results for camera-radar plane transformation. (b) Results for different tracking lengths.}
  \label{fig:trans_testing}
\end{figure}

\subsection{Impact of Different Tracking Length}
Analyzing the impact of tracking or observation length on the performance of MMW-Carry provides insights into optimizing its functionality. The evaluation of the system with knwlTrf across different tracking lengths, as illustrated in Fig.~\ref{fig:trans_testing}(b), reveals a discernible trend. With longer tracking lengths (i.e., an increased number of frames), the accuracy consistently rises, accompanied by a concurrent decrease in both FPR and MR. Specifically, significant performance enhancement is observed as the tracking length extends from 1 to 150 frames. During this range, accuracy ascends from approximately 60\% to 80\%, while FPR and MR decrease from around 40\% to 20\%. However, beyond a tracking length of 150 frames, minimal changes occur in all three metrics. This suggests that an observation length of 150 frames strikes an optimal balance between performance and the practical consideration of the required number of frames for tracking walking individuals.

\subsection{Performance of Different Combination Methods\label{sec:combines}}
In Table~\ref{tab:combine_Res}, we dive into further exploration of the performance gains through different methods of combining multiple observations. Specifically, we compare the proposed knwlTrf with baselines that employ deep learning-based feature fusion: feaCat, faeAtt, feaAcm. We utilized 10 frames as input to constrain the network size. The results in Table~\ref{tab:combine_Res} showcase that the feaCat method attains the lowest FPR. On the other hand, the knwlTrf method achieves the lowest MR, the highest accuracy, and a comparable false alarm rate to feaCat, all without introducing additional layers to the deep learning architecture. This underscores the effectiveness of the knwlTrf approach in achieving a favorable trade-off between detection performance metrics.
\begin{table}
    \centering
    \setlength\tabcolsep{3pt}
     \caption{Comparison between different combination ways of using 10 frames of data.}
    \begin{tabular}{cccc}
        \hline
         Methods & FPR & MR & Accuracy\\
        \hline
         feaCat & \textbf{0.2912} & 0.2982 & 0.7066 \\
         faeAtt & 0.2979 & 0.2891 & 0.7041\\
         feaAcm & 0.3141 & 0.3097 & 0.6883\\
         knwlTrf (\textbf{ours}) & 0.3030 & \textbf{0.2716} & \textbf{0.7110}\\
         \hline
    \end{tabular}
    \label{tab:combine_Res}
\end{table}

\subsection{Camera-aided vs. Radar-only Tracking}
The implemented camera-aided tracking encompasses human detection, tracking, and camera-to-radar plane transformation, enhancing individual localization in complex indoor environments. To assess its efficacy, we conducted a comparative analysis between the camera-aided tracking and traditional radar-only tracking. The radar-only tracking baseline employs the Kalman Filter on clustered non-static CFAR detection points \cite{gao2021mimosar} to generate tracklets, without utilizing any camera information.

\subsubsection{Metrics} The metrics used for evaluating individual tracking performance were the MR and FPR as well. A miss is recorded when the system fails to detect, track, and localize an individual present in the image, while a false alarm is registered when an undesired human tracklet is present.

\begin{table}
    \centering
    \setlength\tabcolsep{3pt} 
    \caption{Evaluation for tracking method.}
    \begin{tabular}{cccc}
        \hline
        Tracking methods & MR & FPR \\
        \hline
         camera-aided  & 0.0874\% & 0.0238\%\\
         radar-only & 19.78\% & 35.19\%\\
        \hline
    \end{tabular}
    \label{tab:img_tracking_Resl}
\end{table}

\begin{figure}
\centering
\includegraphics[width=0.43\textwidth]{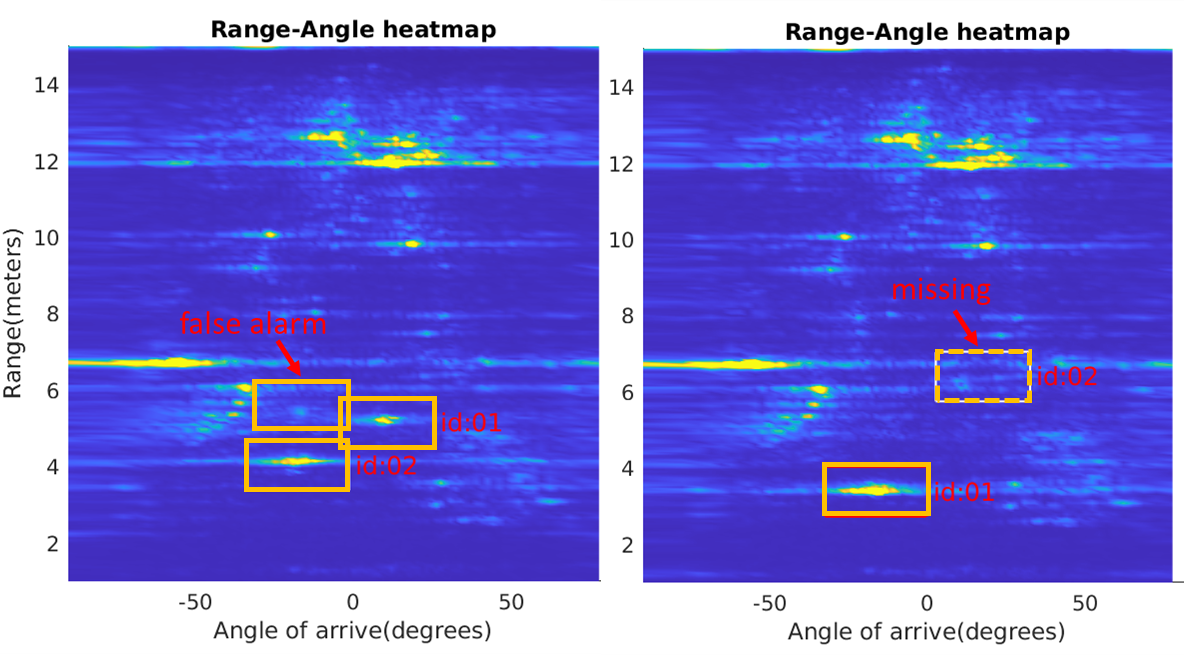}
\caption{Visualizing the common issues of radar-only tracking.}
\label{fig:tracking_comp}
\end{figure}

\subsubsection{Numerical Tracking Performance}
We compared camera-aided tracking to radar-only tracking using the mentioned metrics, and the results are presented in Table~\ref{tab:img_tracking_Resl}. The ground truth was meticulously labeled manually using the camera images. The camera-aided tracking exhibited a significant reduction in MR (from 19.78\% to 0.0874\%) and FPR (from 35.19\% to 0.0238\%). This suggests a reliable identification of human subjects with very few misses and false detections. Visualization of common issues in radar-only tracking is shown in Fig.~\ref{fig:tracking_comp}, where the left part displays frequent false alarms due to multi-path reflections, and the right part shows a missing case typically occurring when individuals are static or close to a strong-reflection background. These issues lead to unexpected tracklets or missing ones during tracking.

\subsection{Discussion}

\begin{table}
    \centering
    \caption{Results for different experiment scenarios.}
    \setlength\tabcolsep{3pt} 
    \begin{tabular}{c|c|ccc}
        \hline
         \multicolumn{2}{c}{Experiment} & FPR & MR & Accuracy\\
        \hline
         \multirow{2}{*}{Occlusion} & Open carry & 0.2000 & 0.1896 & 0.8030\\
         & Concealed & 0.3060 & 0.2534 & 0.7201\\
         \hline
         \multirow{2}{*}{Object Num.}  &
         one & 0.2211 & 0.2028 & 0.7887\\
         & multiple & 0.4941 & 0.2787 & 0.6199\\
        \hline
         \multirow{2}{*}{Individual Num.} & one & 0.2211 & 0.2028 & 0.7887\\
         & multiple & 0.3005 & 0.2972 & 0.6953\\
         \hline
    \end{tabular}
    \label{tab:occlusion_Resl}
\end{table}
\subsubsection{Open Carry vs. Concealed}
One of the primary advantages of the MMW-Carry system is its ability to detect concealed objects with reduced intrusiveness. We evaluated the performance for open carry and concealed objects separately using the collected dataset, and the results are presented in Table~\ref{tab:occlusion_Resl}. In comparison to open carry, the FPR for concealed objects increased from 20\% to 30.6\%, and the MR increased from 18.96\% to 25.34\%. The presence of barriers such as clothes and backpacks between the testbed and objects of interest affects the strengths of the radar signal and introduces additional reflections. Therefore, a reduction in performance is expected when objects are concealed.

\subsubsection{Different Object Amounts}
Detecting multiple small objects on a single test subject is challenging due to the objects being in close proximity to each other, causing their reflection signals to mix. To examine this challenge, we compared the system's performance for a single person carrying one or multiple objects, as shown in Table \ref{tab:occlusion_Resl}. All three performance metrics are noticeably affected when the test subject carries multiple objects. The FPR is the most affected metric, doubling from 22.11\% to 49.41\%. This increase is mainly attributed to the incorrect detection of small objects, such as phones, when limited data of multiple objects exist in the training set, making it challenging for the system to identify the mixed reflection signals. The impact on MR and accuracy is noticeable but less severe, with the MR increasing from 20.28\% to 27.87\% and accuracy reducing from 78.87\% to 61.99\%.

\subsubsection{Single vs. Multiple Individuals Scenario}
The system was initially trained using data from a single moving subject. To assess its generality, we tested the trained system on newly collected data featuring multiple walking subjects. The results, presented in Table \ref{tab:occlusion_Resl}, indicate that the system performs well in these ``unseen" multiple-people scenarios, with only a minor performance drop of approximately 0.1 for all three metrics. This drop is likely attributable to between-subject interference, such as occlusion from other people, leading to missing or altered reflection patterns.

% \subsubsection{Impact of Radar Parameters}
% \highlighttext{The selection of radar parameters significantly impacts the resolution of the radar image formed. Of particular interest are two resolutions: range resolution and angular resolution. Range resolution, defined by $\frac{c}{2B}$, where $B$ represents the sweeping bandwidth, is crucial. A wider bandwidth results in finer range resolution. In this study, we opted for a \SI{2.5}{GHz} bandwidth to strike a balance between range resolution and sweeping time efficiency, considering the available $77$-$\SI{81}{GHz}$ radar band. Angular resolution, on the other hand, relies on the array aperture size. A larger number of antennas yields a larger aperture size and better angular resolution, but practical constraints often limit it. Our radar configuration, equipped with 12 TX and 16 RX antennas, boasts a relatively large array aperture. Additionally, by employing TDM on TXs, we effectively enlarge the virtual antenna aperture size through a MIMO virtual array setup.}

\begin{figure}[t]
\centering
  \includegraphics[width=.49\textwidth]{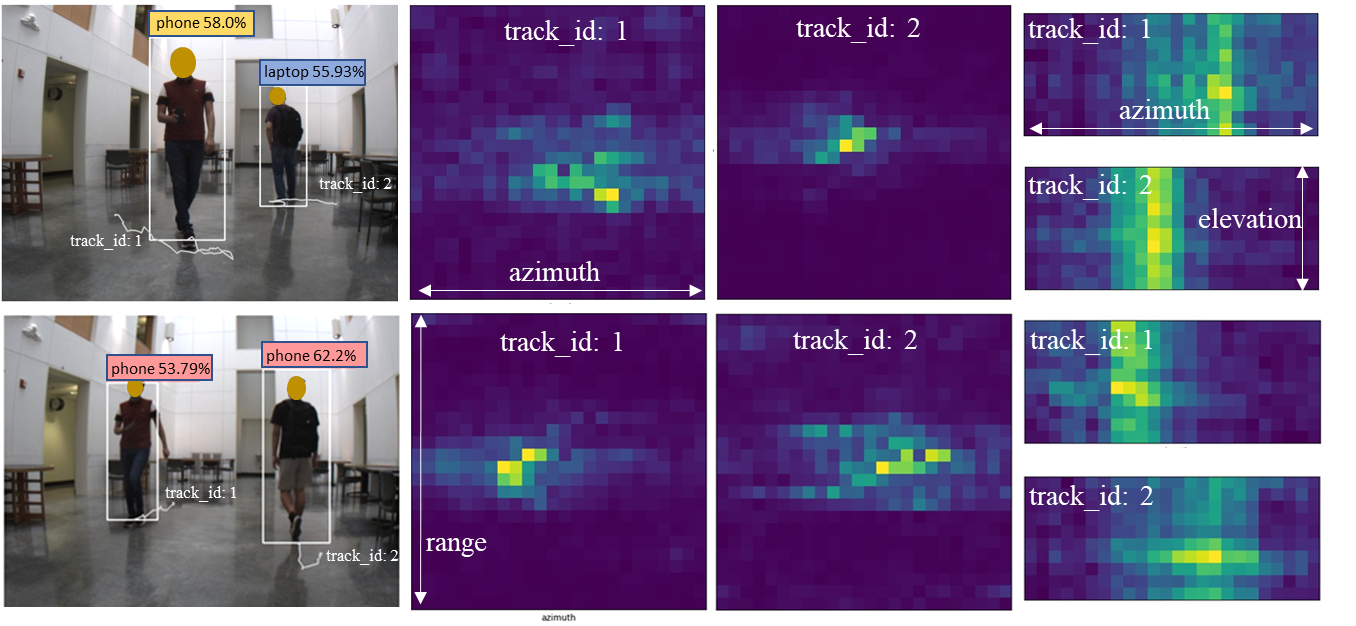}
  \caption{Qualitative results of two samples with the RGB image (tracking trajectories and human detection) and radar image visualization (range-azimuth angle image and elevation-azimuth angle image) for each individual. 
  % Details of 2 samples are explained in the text.
  }
  \label{fig:res_qual}
\end{figure}

\subsection{Qualitative Results}
Fig.~\ref{fig:res_qual} showcases the active performance of our system through four test samples. Each sample includes visualizations of the input cube for each tracked individual in range-azimuth and elevation-azimuth formats. Additionally, RGB images display the individual's trajectory alongside object detection results. These examples illustrate various scenarios, such as 1) Two individuals walking with an open-carry knife and an in-bag laptop; 2) One individual walking with an open-carry knife and an in-pocket phone; 3) Two individuals walking with an open-carry knife and an in-bag knife separately; and 4) One individual rotating with an in-bag knife and laptop. The diversity in object placement, whether in a pocket, backpack, held to the side, or in front, is evident, as well as randomized walking patterns. Despite these varied situations, our system adeptly detects objects by outputting probabilities exceeding the designated threshold. This adaptability underscores the robustness of our approach.

% \subsection{Running Time Analysis}
% We conducted an analysis of the average processing time for our system running on a TITAN RTX GPU. With preprocessed radar images, the system achieves an average running time of \SI{12.53}{ms} per frame, equivalent to approximately \SI{80}{fps}, owing to the lightweight design of our DL model. This demonstrates the feasibility of real-time detection using GPU acceleration. Moreover, our knwlTrf post-processing module supports predictions for each timeslot without the need for input or feature caching, resulting in minimal additional processing time and memory consumption. In comparison with state-of-the-art methods that utilize high-resolution radiographic images, requiring lengthy scanning times (tens of seconds) for generation, our system takes simple 3D radar images as input, delivering rapid processing and acceptable performance.

\section{Conclusion}
% This paper introduces MMW-Carry to predict the existence of objects carried by individuals using MMWave radar signals with cameras. Compared to state-of-the-art methods, it incorporates cameras and long-term multi-observation combinations to enhance the system's robustness and accuracy. Thus, MMW-Carry represents a significant advancement in RF-based sensing, opening up possibilities for applications in safety and smart homes. For future work, we intend to replace FFT with the sparse and super-solution DoA algorithms for further improved imaging resolution and reduced sidelobes.
\highlighttext{In conclusion, our paper presents MMW-Carry, a system combining MMWave radar and visual camera modalities for efficient object detection on moving subjects. Our approach offers a fast, cost-effective, and less intrusive solution. We introduce a method integrating camera-based detection, tracking, and cross-sensor plane transformation to accurately localize individuals in challenging environments. Additionally, we propose the knwlTrf late-fusion module, improving system performance without adding DL layers. Extensive experiments validate our system's effectiveness across diverse scenarios, demonstrating its potential for real-world applications. }

\balance
\bibliographystyle{IEEEtran}
\bibliography{bibtex}

% Generated by IEEEtran.bst, version: 1.14 (2015/08/26)
\begin{thebibliography}{10}
\providecommand{\url}[1]{#1}
\csname url@samestyle\endcsname
\providecommand{\newblock}{\relax}
\providecommand{\bibinfo}[2]{#2}
\providecommand{\BIBentrySTDinterwordspacing}{\spaceskip=0pt\relax}
\providecommand{\BIBentryALTinterwordstretchfactor}{4}
\providecommand{\BIBentryALTinterwordspacing}{\spaceskip=\fontdimen2\font plus
\BIBentryALTinterwordstretchfactor\fontdimen3\font minus
  \fontdimen4\font\relax}
\providecommand{\BIBforeignlanguage}[2]{{%
\expandafter\ifx\csname l@#1\endcsname\relax
\typeout{** WARNING: IEEEtran.bst: No hyphenation pattern has been}%
\typeout{** loaded for the language `#1'. Using the pattern for}%
\typeout{** the default language instead.}%
\else
\language=\csname l@#1\endcsname
\fi
#2}}
\providecommand{\BIBdecl}{\relax}
\BIBdecl

\bibitem{1406480}
H.-M. Chen, S.~Lee, R.~Rao, M.-A. Slamani, and P.~Varshney, ``Imaging for
  concealed weapon detection: a tutorial overview of development in imaging
  sensors and processing,'' \emph{IEEE Signal Processing Magazine}, vol.~22,
  no.~2, pp. 52--61, 2005.

\bibitem{TSA}
\BIBentryALTinterwordspacing
TSA. (2022) Imaging technology. [Online]. Available:
  \url{https://web.archive.org/web/20100106043039/http://www.tsa.gov/approach/tech/imaging_technology.shtm}
\BIBentrySTDinterwordspacing

\bibitem{Roomi2012DETECTIONOC}
D.~M.~M. Roomi and R.Rajashankari, ``\BIBforeignlanguage{English}{Detection of
  concealed weapons in x-ray images using fuzzy k-nn},''
  \emph{\BIBforeignlanguage{English}{International Journal of Computer Science,
  Engineering and Information Technology (IJCSEIT)}}, vol.~2, no.~2, pp.
  187--196, Apr. 2012.

\bibitem{817171}
P.~Varshney, H.-M. Chen, L.~Ramac, M.~Uner, D.~Ferris, and M.~Alford,
  ``Registration and fusion of infrared and millimeter wave images for
  concealed weapon detection,'' in \emph{Proceedings 1999 International
  Conference on Image Processing (Cat. 99CH36348)}, vol.~3, 1999, pp. 532--536
  vol.3.

\bibitem{4682606}
K.~B. Cooper, R.~J. Dengler, N.~Llombart, T.~Bryllert, G.~Chattopadhyay,
  E.~Schlecht, J.~Gill, C.~Lee, A.~Skalare, I.~Mehdi, and P.~H. Siegel,
  ``Penetrating 3-d imaging at 4- and 25-m range using a submillimeter-wave
  radar,'' \emph{IEEE Transactions on Microwave Theory and Techniques},
  vol.~56, no.~12, pp. 2771--2778, 2008.

\bibitem{identf2012}
P.~S.~K. Bandyopadhyay, B.~Datta, and S.~Roy, ``Identifications of concealed
  weapon in a human body,'' 2012, available at
  \url{https://arxiv.org/ftp/arxiv/papers/1210/1210.5653.pdf}.

\bibitem{ramp}
X.~Gao, G.~Xing, S.~Roy, and H.~Liu, ``Ramp-cnn: A novel neural network for
  enhanced automotive radar object recognition,'' \emph{IEEE Sensors Journal},
  vol.~21, no.~4, pp. 5119--5132, 2021.

\bibitem{gao2022learning}
X.~Gao, H.~Liu, S.~Roy, G.~Xing, A.~Alansari, and Y.~Luo, ``Learning to detect
  open carry and concealed object with 77 ghz radar,'' \emph{IEEE journal of
  selected topics in signal processing}, vol.~16, no.~4, pp. 791--803, 2022.

\bibitem{5666180}
S.~Yeom, D.-S. Lee, J.-Y. Son, and S.-H. Kim, ``Concealed object detection
  using passive millimeter wave imaging,'' in \emph{2010 4th International
  Universal Communication Symposium}, 2010, pp. 383--386.

\bibitem{11198926}
J.~Liu, K.~Zhang, Z.~Sun, Q.~Wu, W.~He, and H.~Wang, ``Concealed object
  detection and recognition system based on millimeter wave fmcw radar,''
  \emph{Applied Sciences}, vol.~11, no.~19, 2021.

\bibitem{5530374}
X.~Zhuge and A.~G. Yarovoy, ``A sparse aperture mimo-sar-based uwb imaging
  system for concealed weapon detection,'' \emph{IEEE Transactions on
  Geoscience and Remote Sensing}, vol.~49, no.~1, pp. 509--518, 2011.

\bibitem{8650148}
X.~Wang, S.~Gou, X.~Wang, Y.~Zhao, and L.~Zhang, ``Patch-based gaussian mixture
  model for concealed object detection in millimeter-wave images,'' in
  \emph{TENCON 2018 - 2018 IEEE Region 10 Conference}, 2018, pp. 2522--2527.

\bibitem{10.1145/3447993.3483258}
\BIBentryALTinterwordspacing
T.~Zheng, Z.~Chen, J.~Luo, L.~Ke, C.~Zhao, and Y.~Yang, ``Siwa: See into walls
  via deep uwb radar,'' in \emph{Proceedings of the 27th Annual International
  Conference on Mobile Computing and Networking}, ser. MobiCom '21.\hskip 1em
  plus 0.5em minus 0.4em\relax New York, NY, USA: Association for Computing
  Machinery, 2021, p. 323–336. [Online]. Available:
  \url{https://doi.org/10.1145/3447993.3483258}
\BIBentrySTDinterwordspacing

\bibitem{gao2021mimosar}
X.~Gao, S.~Roy, and G.~Xing, ``Mimo-sar: A hierarchical high-resolution imaging
  algorithm for mmwave fmcw radar in autonomous driving,'' \emph{IEEE
  Transactions on Vehicular Technology}, vol.~70, no.~8, pp. 7322--7334, 2021.

\bibitem{gao2021perception}
X.~Gao, S.~Roy, G.~Xing, and S.~Jin, ``Perception through 2d-mimo fmcw
  automotive radar under adverse weather,'' in \emph{2021 IEEE International
  Conference on Autonomous Systems (ICAS)}, 2021, pp. 1--5.

\bibitem{gao2019experiments}
X.~{Gao}, G.~{Xing}, S.~{Roy}, and H.~{Liu}, ``Experiments with mmwave
  automotive radar test-bed,'' in \emph{2019 53rd Asilomar Conference on
  Signals, Systems, and Computers}, 2019, pp. 1--6.

\bibitem{ti_mimo}
\BIBentryALTinterwordspacing
S.~Rao, \emph{White paper: MIMO Radar}.\hskip 1em plus 0.5em minus 0.4em\relax
  Texas Instrument, 2017, no. SWRA554A. [Online]. Available:
  \url{https://www.ti.com/lit/an/swra554a}
\BIBentrySTDinterwordspacing

\bibitem{8519763}
Y.~Li, Z.~Peng, R.~Pal, and C.~Li, ``Potential active shooter detection based
  on radar micro-doppler and range-doppler analysis using artificial neural
  network,'' \emph{IEEE Sensors Journal}, vol.~19, no.~3, pp. 1052--1063, 2019.

\bibitem{8536660}
Z.~Zhang, X.~Di, Y.~Xu, and J.~Tian, ``Concealed dangerous object detection
  based on a 77ghz radar,'' in \emph{2018 IEEE International Workshop on
  Electromagnetics:Applications and Student Innovation Competition (iWEM)},
  2018, pp. 1--2.

\bibitem{10188595}
A.~Sonny, A.~Kumar, and L.~R. Cenkeramaddi, ``Carry object detection utilizing
  mmwave radar sensors and ensemble-based extra tree classifiers on the edge
  computing systems,'' \emph{IEEE Sensors Journal}, vol.~23, no.~17, pp.
  20\,137--20\,149, 2023.

\bibitem{8628238}
T.~Liu, Y.~Zhao, Y.~Wei, Y.~Zhao, and S.~Wei, ``Concealed object detection for
  activate millimeter wave image,'' \emph{IEEE Transactions on Industrial
  Electronics}, vol.~66, no.~12, pp. 9909--9917, 2019.

\bibitem{9269991}
C.~Wang, J.~Shi, Z.~Zhou, L.~Li, Y.~Zhou, and X.~Yang, ``Concealed object
  detection for millimeter-wave images with normalized accumulation map,''
  \emph{IEEE Sensors Journal}, vol.~21, no.~5, pp. 6468--6475, 2021.

\bibitem{9353483}
M.~T. Bhatti, M.~G. Khan, M.~Aslam, and M.~J. Fiaz, ``Weapon detection in
  real-time cctv videos using deep learning,'' \emph{IEEE Access}, vol.~9, pp.
  34\,366--34\,382, 2021.

\bibitem{9211448}
O.~Bazgir, D.~Nolte, S.~R. Dhruba, Y.~Li, C.~Li, S.~Ghosh, and R.~Pal, ``Active
  shooter detection in multiple-person scenario using rf-based machine
  vision,'' \emph{IEEE Sensors Journal}, vol.~21, no.~3, pp. 3609--3622, 2021.

\bibitem{equipment2018user}
3GPP and 38.101, ``5g; nr; user equipment (ue) radio transmission and
  reception,'' \emph{Part 2: Range 2 Standalone, 2 Version 15.2.0 Release 15},
  2018.

\bibitem{richards2014fundamentals}
M.~A. Richards, \emph{Fundamentals of radar signal processing}.\hskip 1em plus
  0.5em minus 0.4em\relax McGraw-Hill Education, 2014.

\bibitem{nitzberg1972constant}
R.~Nitzberg, ``Constant-false-alarm-rate signal processors for several types of
  interference,'' \emph{IEEE Transactions on Aerospace and Electronic Systems},
  no.~1, pp. 27--34, 1972.

\bibitem{5509644}
A.~Coates and A.~Y. Ng, ``Multi-camera object detection for robotics,'' in
  \emph{2010 IEEE International Conference on Robotics and Automation}, 2010,
  pp. 412--419.

\bibitem{ti_casd}
T.~Instrument, \emph{White paper: Imaging Radar Using Cascaded mmWave Sensor
  Reference Design}.\hskip 1em plus 0.5em minus 0.4em\relax Texas Instrument,
  2019, no. TIDUEN5A, available at
  \url{https://www.ti.com/lit/ug/tiduen5a/tiduen5a.pdf}.

\bibitem{gao2023static}
X.~Gao, S.~Roy, and L.~Zhang, ``Static background removal in vehicular radar:
  Filtering in azimuth-elevation-doppler domain,'' \emph{arXiv preprint
  arXiv:2307.01444}, 2023.

\bibitem{526899}
H.~{Krim} and M.~{Viberg}, ``Two decades of array signal processing research:
  the parametric approach,'' \emph{IEEE Signal Processing Magazine}, vol.~13,
  no.~4, pp. 67--94, 1996.

\bibitem{CHUNG2014599}
\BIBentryALTinterwordspacing
P.-J. Chung, M.~Viberg, and J.~Yu, ``Chapter 14 - doa estimation methods and
  algorithms,'' in \emph{Academic Press Library in Signal Processing: Volume
  3}, ser. Academic Press Library in Signal Processing, A.~M. Zoubir,
  M.~Viberg, R.~Chellappa, and S.~Theodoridis, Eds.\hskip 1em plus 0.5em minus
  0.4em\relax Elsevier, 2014, vol.~3, pp. 599--650. [Online]. Available:
  \url{https://www.sciencedirect.com/science/article/pii/B978012411597200014X}
\BIBentrySTDinterwordspacing

\bibitem{190331}
J.~McWirter and T.~Shepherd, ``A systolic array processor for mvdr
  beamforming,'' in \emph{IEE Colloquium on Adaptive Antennas}, 1990, pp.
  4/1--4/2.

\bibitem{1143830}
R.~Schmidt, ``Multiple emitter location and signal parameter estimation,''
  \emph{IEEE Transactions on Antennas and Propagation}, vol.~34, no.~3, pp.
  276--280, 1986.

\bibitem{lsvm-pami}
P.~F. Felzenszwalb, R.~B. Girshick, D.~McAllester, and D.~Ramanan, ``Object
  detection with discriminatively trained part based models,'' \emph{IEEE
  Transactions on Pattern Analysis and Machine Intelligence}, vol.~32, no.~9,
  pp. 1627--1645, 2010.

\bibitem{kalman}
\BIBentryALTinterwordspacing
R.~E. Kalman, ``{A New Approach to Linear Filtering and Prediction Problems},''
  \emph{Journal of Basic Engineering}, vol.~82, no.~1, pp. 35--45, 03 1960.
  [Online]. Available: \url{https://doi.org/10.1115/1.3662552}
\BIBentrySTDinterwordspacing

\bibitem{9341164}
X.~Weng, J.~Wang, D.~Held, and K.~Kitani, ``3d multi-object tracking: A
  baseline and new evaluation metrics,'' in \emph{2020 IEEE/RSJ International
  Conference on Intelligent Robots and Systems (IROS)}, 2020, pp.
  10\,359--10\,366.

\bibitem{Jonker2005ASA}
R.~Jonker and A.~Volgenant, ``A shortest augmenting path algorithm for dense
  and sparse linear assignment problems,'' \emph{Computing}, vol.~38, pp.
  325--340, 2005.

\bibitem{1334537}
S.~Sugimoto, H.~Tateda, H.~Takahashi, and M.~Okutomi, ``Obstacle detection
  using millimeter-wave radar and its visualization on image sequence,'' in
  \emph{Proceedings of the 17th International Conference on Pattern
  Recognition, 2004. ICPR 2004.}, vol.~3, 2004, pp. 342--345 Vol.3.

\bibitem{8581329}
J.~Oh, K.-S. Kim, M.~Park, and S.~Kim, ``A comparative study on camera-radar
  calibration methods,'' in \emph{2018 15th International Conference on
  Control, Automation, Robotics and Vision (ICARCV)}, 2018, pp. 1057--1062.

\bibitem{li2007mimo}
J.~Li and P.~Stoica, ``Mimo radar with colocated antennas,'' \emph{IEEE Signal
  Processing Magazine}, vol.~24, no.~5, pp. 106--114, 2007.

\bibitem{doi:10.1036/0071444742}
\BIBentryALTinterwordspacing
M.~A. Richards, ``Fundamentals of radar signal processing.''\hskip 1em plus
  0.5em minus 0.4em\relax US: McGraw-Hill Professional, 2005, pp. --1.
  [Online]. Available:
  \url{https://mhebooklibrary.com/doi/book/10.1036/0071444742}
\BIBentrySTDinterwordspacing

\bibitem{7780459}
K.~He, X.~Zhang, S.~Ren, and J.~Sun, ``Deep residual learning for image
  recognition,'' in \emph{2016 IEEE Conference on Computer Vision and Pattern
  Recognition (CVPR)}, 2016, pp. 770--778.

\bibitem{https://doi.org/10.48550/arxiv.2005.12872}
\BIBentryALTinterwordspacing
N.~Carion, F.~Massa, G.~Synnaeve, N.~Usunier, A.~Kirillov, and S.~Zagoruyko,
  ``End-to-end object detection with transformers,'' 2020. [Online]. Available:
  \url{https://arxiv.org/abs/2005.12872}
\BIBentrySTDinterwordspacing

\bibitem{begn-ye78-22}
\BIBentryALTinterwordspacing
X.~Gao, S.~Roy, H.~Liu, Y.~Luo, and G.~Xing, ``Raw adc data of 2d-mimo mmwave
  radar for carry object detection,'' 2022. [Online]. Available:
  \url{https://dx.doi.org/10.21227/begn-ye78}
\BIBentrySTDinterwordspacing

\end{thebibliography}

\end{document}